\documentclass[preprint,proceedings]{rmaa} 
 
\usepackage{paralist} 
 
\usepackage{psfrag,color} 
 
\newcommand{\mic}{\hbox{$\mu{\rm m}$}} 
\newcommand{\cms}{\hbox{${\rm cm^{-2}}$}}

\newcommand{\anu}{\hbox{$\alpha_{\nu}$}}

\newcommand{\zq}{\hbox{$z_q$}}

\newcommand{\NEVIII}{\hbox{Ne\,{\sc viii}}}

\newcommand{\Lya}{\hbox{Ly$\alpha$}}

\newcommand{\anuv}{\hbox{$\alpha_{NUV}$}} 
\newcommand{\anuva}{\hbox{$\bar \alpha_{NUV}$}} 
\newcommand{\fdi}{\hbox{$f_{D1}$}} 

\newcommand{\NN}{\hbox{$N_{20}$}}

\newcommand{\lar}{\hbox{$\lambda_{rest}$}}

\SetYear{2007} 
\SetConfTitle{Triggering Relativistic Jets} 
 
\title{Dust and the far-UV spectral energy distribution of quasars}  
 
\author{ 
  Luc Binette,\altaffilmark{1}  
  Yair Krongold,\altaffilmark{1}
  Gladis Magris Crestini\altaffilmark{2}  
  and Jose Antonio de Diego\altaffilmark{1}} 
 
\altaffiltext{1}{Instituto de Astronom\'\i{}a, UNAM, M\'exico D.F.,  
M\'exico.} 
\altaffiltext{2}{Centro de Investigaciones de Astronom\'\i a, M\'erida, Venezuela.} 
 
\shortauthor{Binette et al.} 
\shorttitle{Dust and the spectral energy distribution of quasars} 
 
\fulladdresses{ 
\item  Luc Binette, Yair Krongold and Jose Antonio de Diego: Instituto de  
Astronom\'\i{}a, Universidad Nacional 
  Aut\'onoma de M\'exico, Ciudad Universitaria, Apartado Postal 70--264, CP 
  04510, M\'exico D.F., M\'exico (\email{LBinette@astroscu.unam.mx}). 
\item Gladis Magris Crestini: Centro de Investigaciones de Astronom\'\i a, Apartado postal 264, M\'erida 5101-A, Venezuela 
} 
 
\listofauthors{L. Binette, Y. Krongold, G. Magris C., J. A. de Diego} 
\indexauthor{ Binette, L.} 
\indexauthor{Krongold, Y.} 
\indexauthor{Magris C., G.} 
\indexauthor{de Diego, J. A.} 
 
\abstract{The  spectral energy distribution of quasars shows a sharp
steepening of the continuum shortward of  $\simeq 1100$\,\AA. We present a new dust extinction model consisting of crystalline carbon grains. We analyze the unusual extinction properties of this dust and proceed to show how the observed UV  spectral break can be successfully modelled by crystalline carbon dust.}

\resumen{La distribuci\'on de energ\'\i a espectral de los cuas\'ares presenta un empinamiento abrupto del continuo  a longitudes m\'as cortas que $\simeq 1100$\,\AA. 
Presentamos un nuevo modelo de extinci\'on por polvo constituido de  granos de carbono cristalino.
Analizamos las propiedades en extinci\'on poco usuales de este polvo y demostramos como el empinamiento abrupto observado en cu\'asares puede ser modelizado exitosamente por polvo hecho de carbono cristalino. }
 
\addkeyword{ISM: dust } 
\addkeyword{galaxies: active} 
\addkeyword{radiative transfer} 
\addkeyword{ultraviolet: general}

\begin{document} 

\maketitle 
 
\begin{figure*}[!t]
\includegraphics[width=\columnwidth,height=8cm]{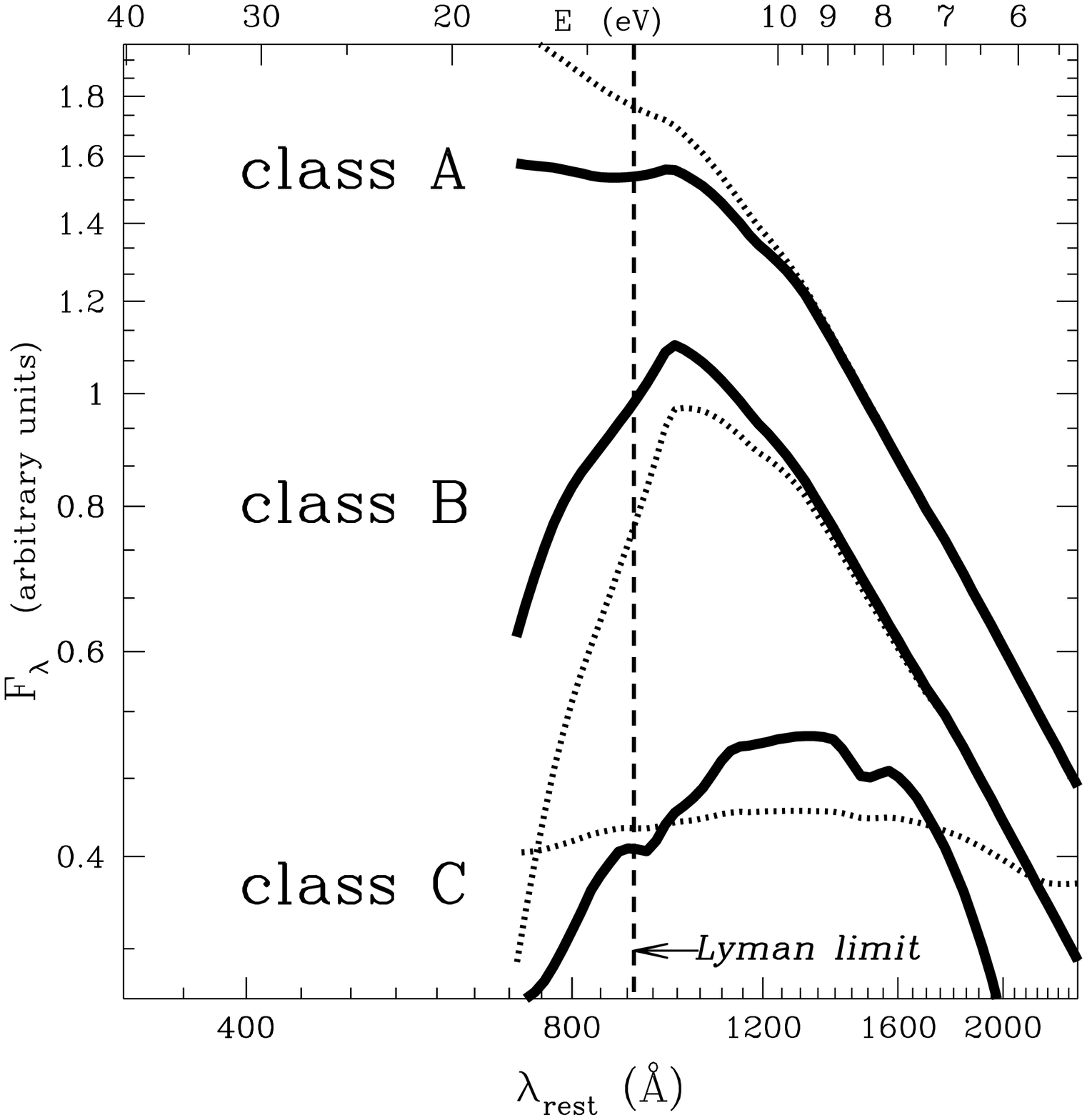}%
\hspace*{\columnsep}%
\includegraphics[width=\columnwidth,height=8cm]{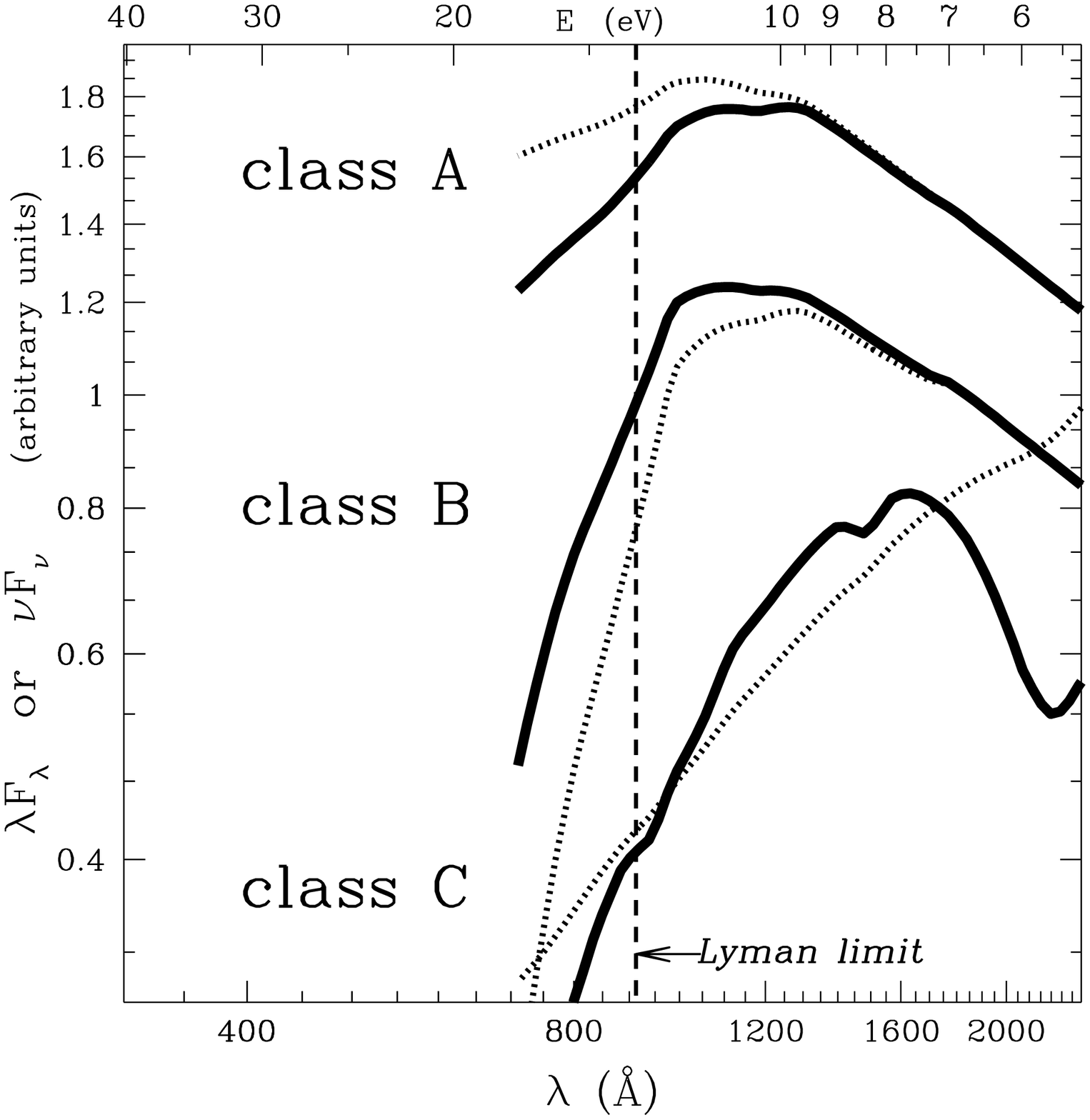}
\caption{Panel a (left):  cartoon that illustrates in
  $F_{\lambda}$ the three main classes of far-UV break (continuous lines) observed among
  the 61 multigrating HST-FOS spectra that extend down to at least
  900\,\AA.  The dotted line  shows typical variations within a
  given class.  Panel b (right): same descriptive cartoon using instead the
  ${\nu}F_{\nu}$ representation. } 
\label{fig:class}
\end{figure*}

\section{Introduction} \label{sec:intro} 
 
The ultraviolet energy distribution of quasars is characterized by the
so-called ``big blue bump'', which peaks in ${\nu}F_{\nu}$ at
approximately 1000\,\AA.  The quasar `composite' spectral energy
distribution (SED) of Telfer et\,al. (2002, hereafter TZ02) obtained
by co-adding 332 HST-FOS archived spectra of 184 quasars between
redshifts 0.33 and 3.6, exhibits a steepening of the continuum at
$\sim 1100$\,\AA. A fit of this composite SED using a broken powerlaw
reveals that the powerlaw index changes from $-0.69$ in the near-UV to
$-1.76$ in the far-UV (see TZ02). We label this observed sharp steepening the
`far-UV break'.  In these proceedings, we first give a short overview of our
classification of the HST-FOS quasar spectra based on the appearance
of the far-UV break. We then  describe our dust absorption model based on
crystalline carbon  grains. We illustrate the extinction
properties of this dust and proceed to compare with two quasar
spectra.  In a previous proceeding (Binette et\,al. 2005b), we
described how we came to consider the possibility of carbon
crystallite dust. The  argumentation in support of the
dust absorption interpretation of the UV break has been fully described 
 in Binette et\,al. (2005a, hereafter BM05). Complementary
information about crystalline dust can be found in this and other proceedings
(Binette et\,al. 2005b, 2006).

\section{Classification of the far-UV break appearance}\label{sec:class}

The concept of a `composite' Spectral Energy Distribution (SED) that
extends into the far-UV rests on the redshift effect, which allows us
to observe the far-UV region of high-$z$quasars, beyond the Lyman
limit, even though our Galaxy is highly opaque in this wavelength
region (observer-frame). For the concept of a `composite' SED obtained
by averaging to work, the environment of quasars must be highly
transparent beyond the Lyman limit, a condition which is generally
satisfied for quasars at $\zq \la 3.6$. In what follows, we
will consider the sample of HST-FOS spectra that H. Telfer kindly
lent to us. These have already been corrected for Galactic
dust extinction, for the presence of the Lyman valley due to unresolved
\Lya\ forest lines and for the presence of Lyman limit systems.  Of
all the spectra available, we consider only those 106 quasars for
which more than one grating was used since otherwise the \lar\
coverage turned out to be too narrow to provide meaningful constraints
to the models. Furthermore, only 61 of these spectra extend sufficiently in
the far-UV (900 \AA) to reveal enough of the far-UV break to be of use
to us.

After reviewing the HST-FOS sample, some patterns emerge among the selected 61
multigrating spectra. The majority of them (44 quasars) resemble the
composite SED of TZ02. We classify those as class (A). Others (6
quasars) show a sharper and much more pronounced break. We
label these class (B). Eight other quasars present a somewhat shallower
break, which appears to start further in the near-UV or to present a SED which is
flat throughout the  $F_{\lambda}$ region covered. A cartoon of theses three classes is presented
in Fig.\,1a.  The dotted lines show typical variations within each
class.  Finally, only 3 quasars did not fit any meaningful pattern
(class D).  In Fig.\,1b, we show the same cartoon, but in
${\nu}F_{\nu}$ ($= {\lambda}F_{\lambda}$).  We found the traditional
$F_{\lambda}$ representation of Fib.\,1a.  to be more appropriate and
useful when fitting the UV break region.

\begin{figure}[!t]
  \includegraphics[width=\columnwidth]{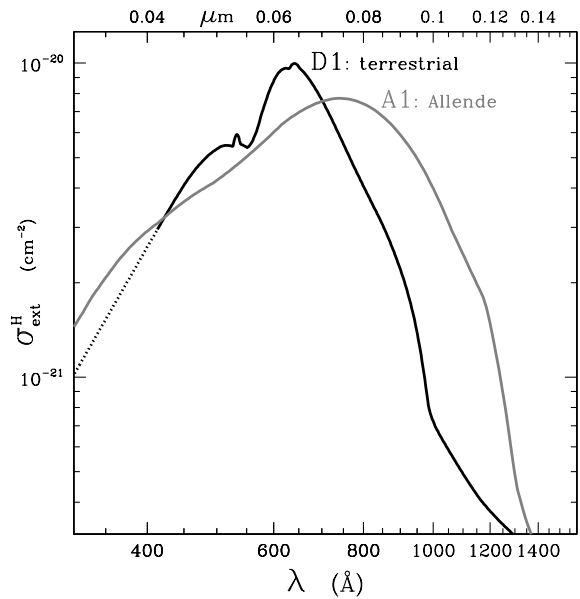}
  \caption{Extinction cross section for nanodiamonds with 
and without  surface impurities (A1 and D1, respectively). The small grain size regime was assumed.
The  complex refraction indices $n+ik$ used in the derivation of the extinction curves are from 
Mutschke et\,al. (2004) for the Allende meteorite nanodiamonds and from
Edwards \& Philipp (1985)  for the cubic
(terrestrial) nanodiamonds.  }
\label{fig:exti}
\end{figure}

\section{The extinction curve produced by crystalline carbon}\label{sec:ext}

At the onset of our project, it became clear that ISM or even SMC-like
dust could not reproduce a break as sharp as the one observed in the
far-UV of quasars. Such a conclusion had been reached by Shang
et\,al. (2005).  A bibliographical research indicated that crystalline
carbon might  possess the required optical properties
(Mutschke et\,al. 2004).  Assuming spherical grains and the Mie
theory, we used the subroutine BHMIE (Bohren and Huffman 1983) to
compute the extinction curves, adopting the complex refraction indices $n+ik$ 
as tabulated in Mutschke et\,al. (2004) and in Edwards \& Philipp (1985) . 
Crystalline carbon can  appear either in
a pure form or with surface impurities. In the former case, they are
referred to as terrestrial cubic diamonds. In the latter case,
nanodiamonds of the type that are  found in abundance in primitive meteorites will
represent the case of crystalline C with impurities. Convincing  evidence of nanodiamonds
being detected in emission is provided by the 3.43 and 3.53 \mic\
emission bands observed in two Ae/Be stars and one post-AGB star (see Van
Kerckhoven, Tielens \& Waelkens 2002). These bands result from the
excitation of surface CH stretching modes (implying the existence of nanodiamonds with H surface
impurities). We present in Fig.\,\ref{fig:exti} the extinction curves
A1 and D1 resulting from nanodiamonds with and without surface
impurities. We assumed a powerlaw grain size distribution within the range 
2 to 25\,\AA.  It turns out that within  this small grain size
regime, using a log-normal distribution or altering slightly the  above size
limits do not produce any differences in the derived extinction
curves, as explained in BM05.  

It is apparent from Fig.\,\ref{fig:exti} that the  steepness of the {\it near-UV}
extinction rise  ought to produce  a sharp absorption break, a desired  feature for the any
dust grain model that aims at fitting the UV quasar break.  The
extinction curves in Fig.\,\ref{fig:exti} were normalized in such a
way that they represent the case of  having all the carbon in the dust (assuming a
solar C/H abundance ratio). This normalization would need to be scaled
according to the actual but  unknown dust-to-gas ratio appropriate
to the quasar environment.

\begin{figure}[!t]
  \includegraphics[width=\columnwidth]{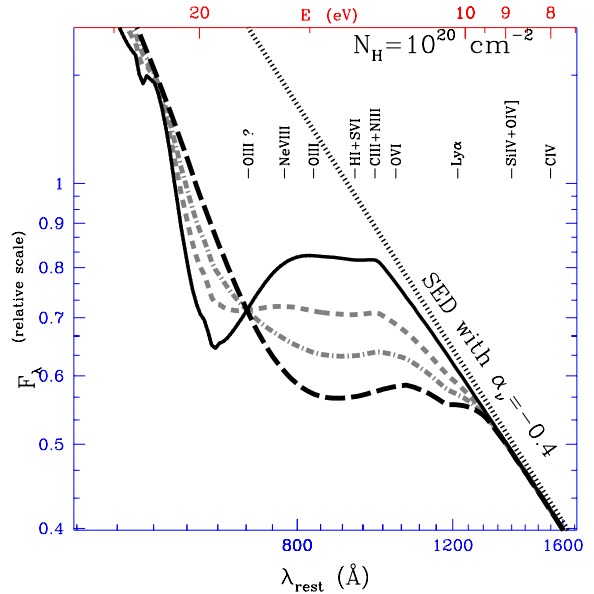}
  \caption{Absorption models assuming a powerlaw SED of index
  $\anu=-0.4$ (dotted line). The nanodiamond dust screen corresponds
  to an hydrogen column of $10^{20}$\,\cms.  The absorbed SED
  represented by the continuous line corresponds to pure cubic
  diamonds (D1) while the long dashed line corresponds to nanodiamonds from
  the Allende meteorite (A1). The grey lines correspond to a mixture of the
  two flavors: dashed line: 60\% D1 + 40\% A1, dot-dashed line: 30\%
  D1 + 70\% A1.  Labelled pointers indicate where emission lines are
  expected in the rest-frame quasar spectra.}
\label{fig:abs}
\end{figure}

\section{Powerlaws absorbed by nanodiamond dust}\label{sec:abs}

Where might the putative crystalline dust be located? In BM05, we
present calculations in which the dust is either intrinsic to the
quasars or intergalactic. In the end, we concluded that only the {\it
intrinsic dust} hypothesis was satisfactory. The calculations
presented here will assume the intrinsic dust case.

\begin{figure}[!t]
  \includegraphics[width=\columnwidth]{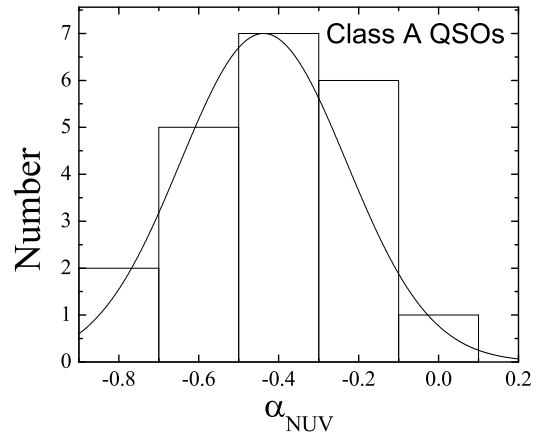}
  \caption{Distribution of the near-UV powerlaw indices found among
  the 21 class (A) quasars for which a reliable estimate of \anuv\ could
  be determined directly from the HST spectra. }
\label{fig:anuv}
\end{figure}

\begin{figure*}[!t]
  \includegraphics[width=\columnwidth,height=8cm]{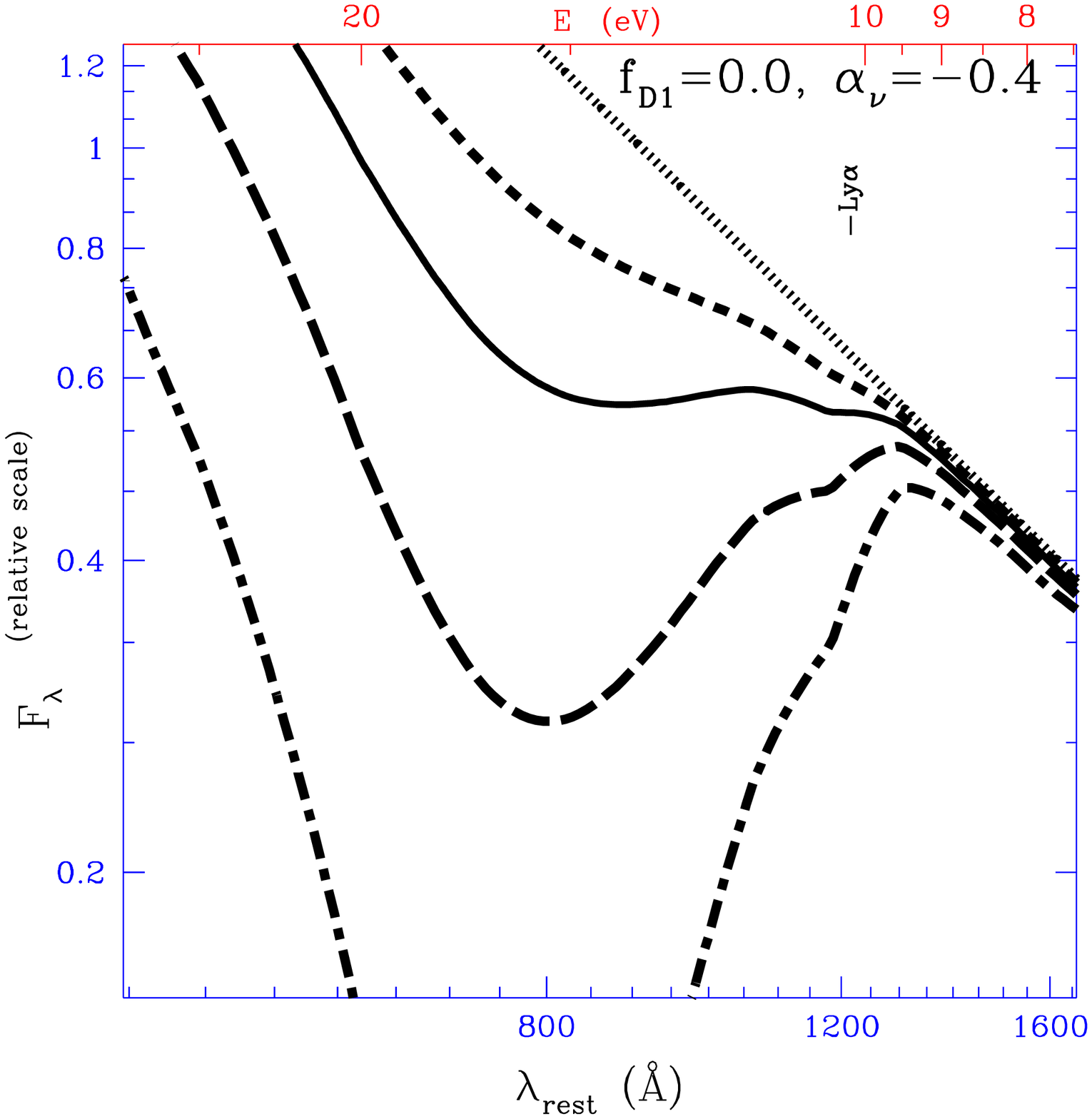}%
  \hspace*{\columnsep}%
  \includegraphics[width=\columnwidth,height=8cm]{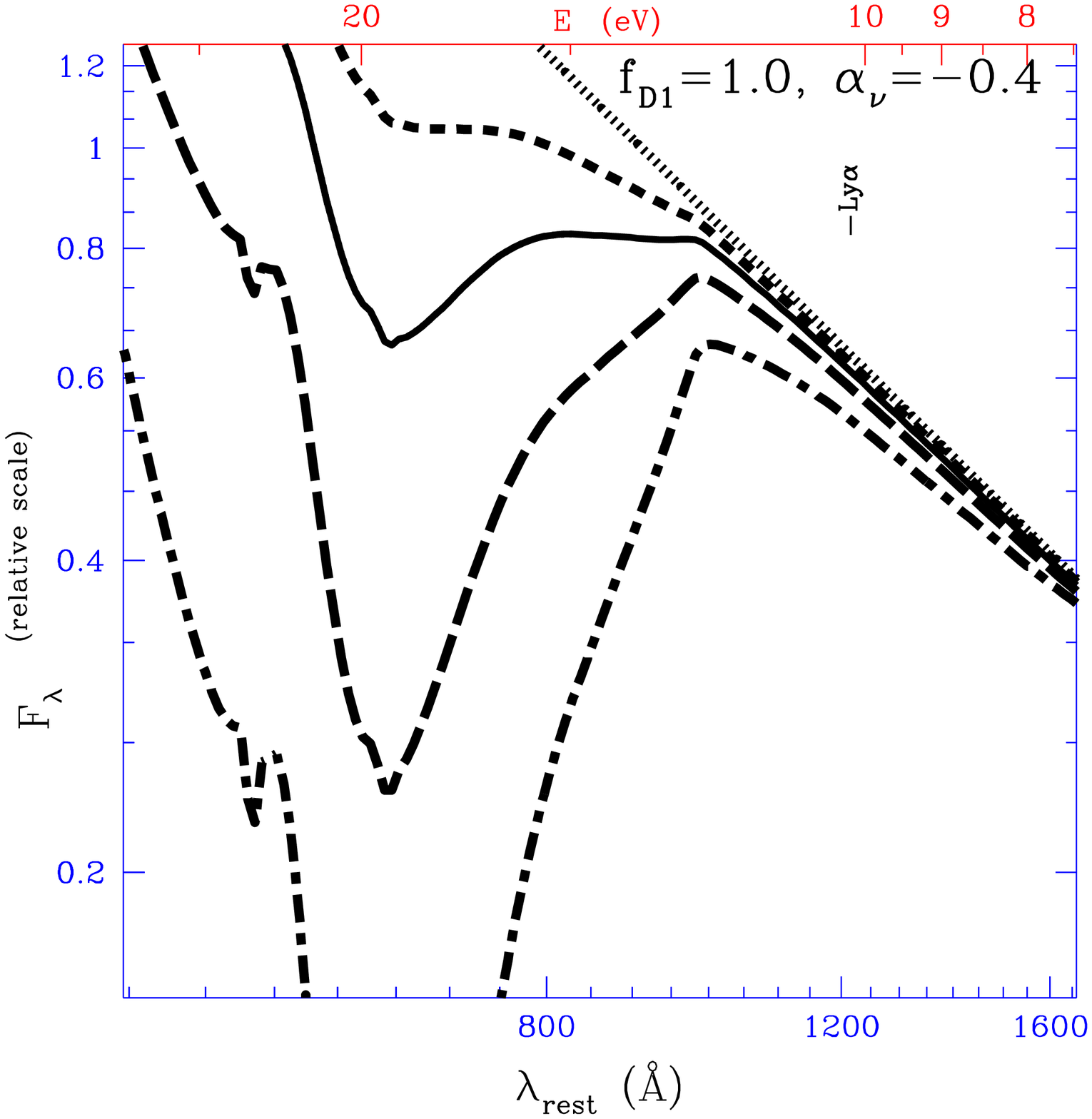}
\caption{Absorbed powerlaw SED assuming a constant index $\anu=-0.4$ (dotted line)
and  the following H columns in units 
of $10^{20}$\,\cms:  $\NN=0.5$, 1.0, 2.0 and 4.0.
Panel a (left):  terrestrial nanodiamond dust (D1). Panel b (right): meteoritic nanodiamonds (A1).
(\fdi\ is the fraction of D1 dust).}
\label{fig:cross}
\end{figure*}

If one fits with a powerlaw the near-UV spectra of class (A)
spectra, one finds that the average spectral index is $\anuva = -0.44$
with a dispersion of 0.21. The corresponding histogram is shown in Fig.\,4. 
This average considers only the 21 objects,
for which a reliable estimate of \anuv\ could be determined directly
from the HST spectra. It is significantly harder than the mean value
of $-0.69$ reported by TZ02 and the median value of $-0.83\pm{0.04}$
for local AGN reported by Scott et\,al. (2004), presumably, because the softer
class (C) spectra are not included in our average.  In
Fig.\,5a and 5b, we present a powerlaw of index $\anu=-0.4$ absorbed
by nanodiamond dust of A1 and D1 types, respectively. The dust screen
takes on four different thicknesses corresponding to H  columns (in
units of $10^{20}$\,\cms) of $\NN=0.5$, 1.0, 2.0 and 4.0. When
$\NN \approx 1.0$, the absorption results in a near horizontal $F_{\lambda}$
spectral segment shortward of the break. The position of the
break occurs at different wavelengths, depending on the dust
composition (compare Fig.\,5a and 5b). It occurs longward and shortward
of \Lya\ in the case of the A1 and D1 dust, respectively. Furthermore,
terrestrial diamond dust can result in a conspicuous absorption dip
near 650\,\AA\ (continuous line in Fig.\,5b and 3), which may have some
relevance to the narrow absorption dip reported by Scott et\,al. in
their composite SED,  which was interpreted as blueshifted \NEVIII\
absorption.

\begin{figure*}[!t]
  \includegraphics[width=\columnwidth,height=8cm]{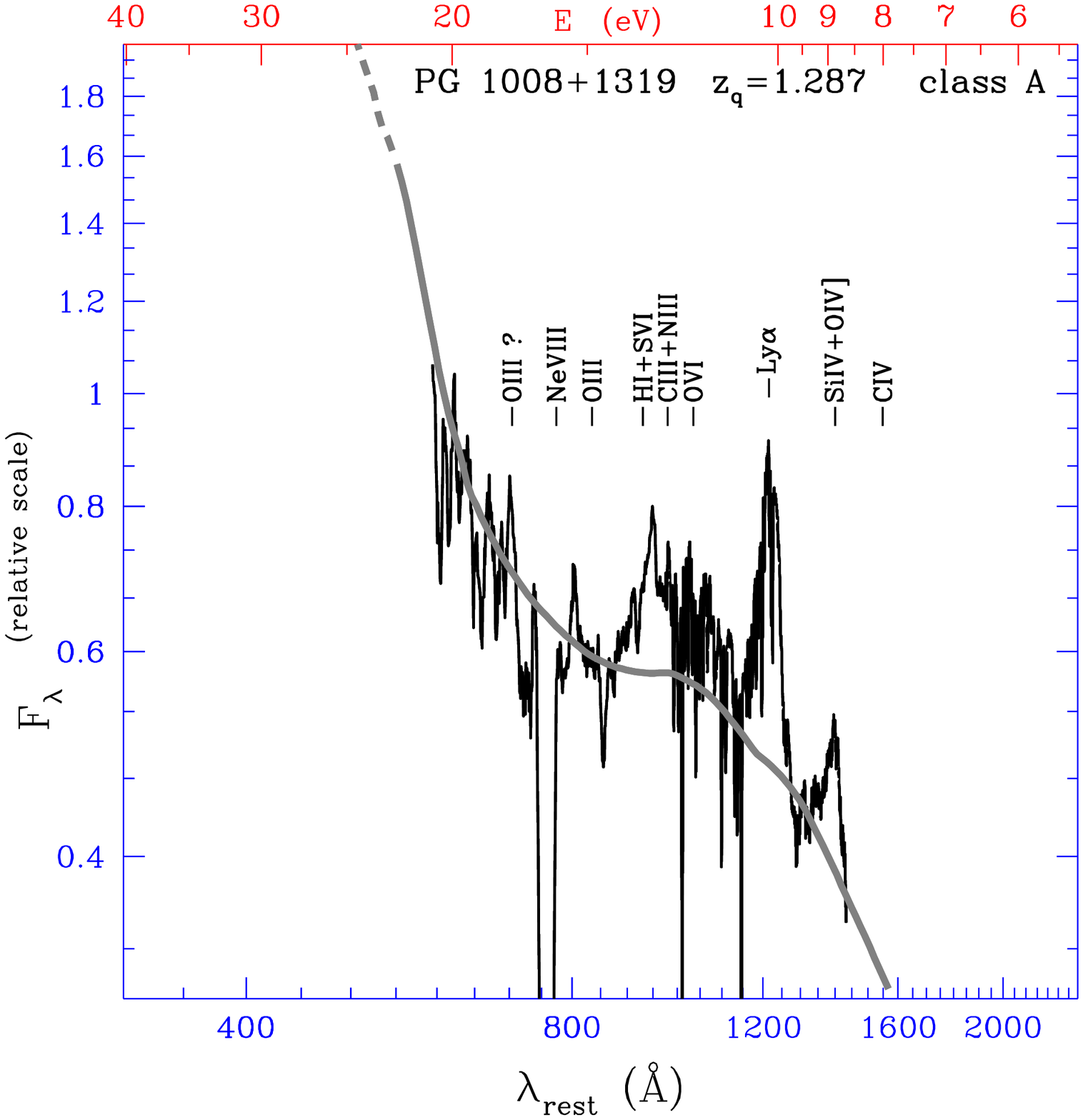}%
  \hspace*{\columnsep}%
  \includegraphics[width=\columnwidth,height=8cm]{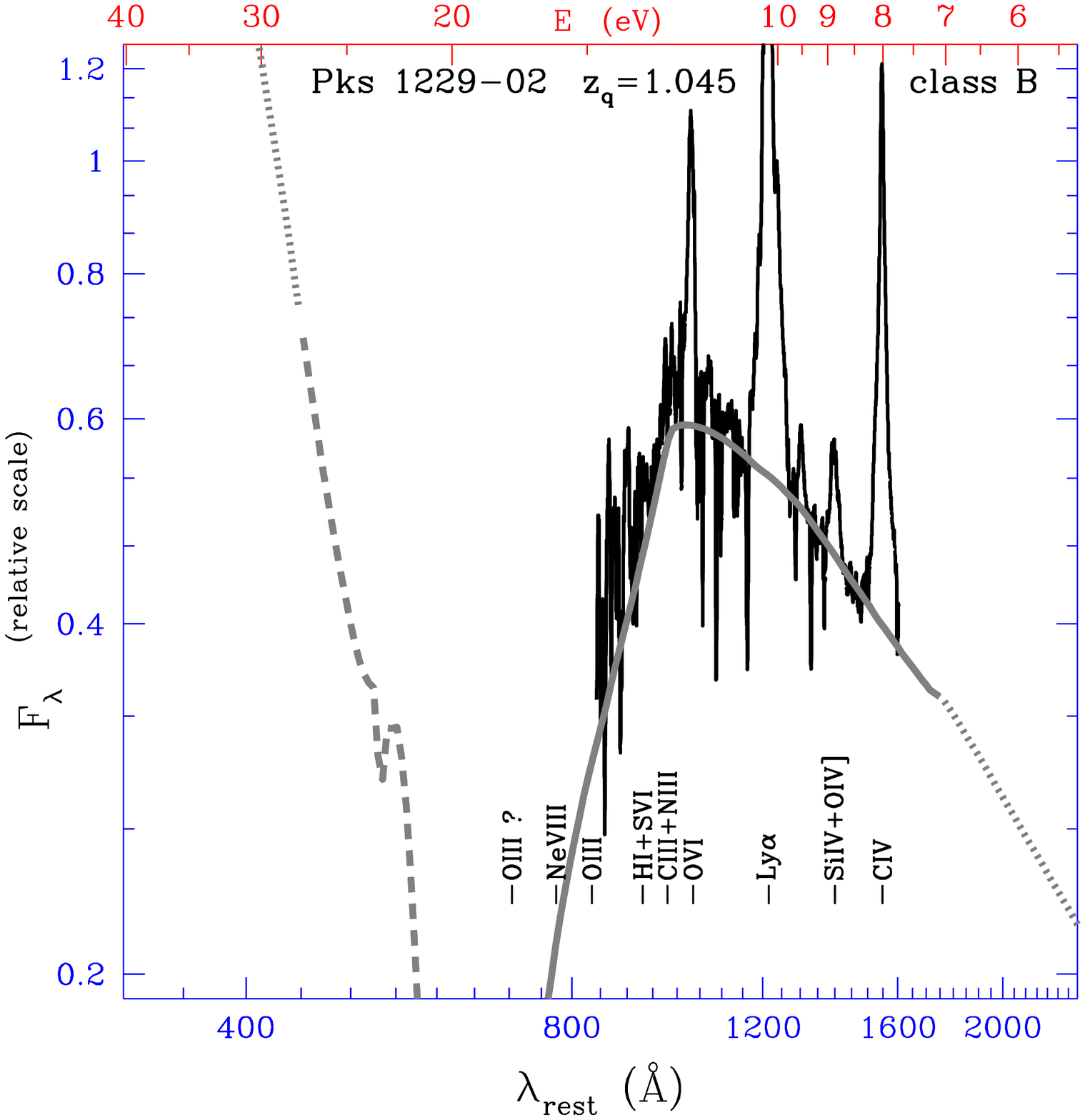}
\caption{Left panel a: rest-frame spectrum of class (A) quasar PG\,1008+1319.  Notice the far-UV rise
shortward of the UV break. The model (grey line) assumes $\anu=+0.13$,
\NN=1.2 and \fdi=0.3. Right panel b: rest-frame spectrum of class (B)
quasar Pks\,1229$-$02. The model (grey line) assumes $\anu=-0.4$,
\NN=3.6 and \fdi=0.90. This is the largest column encountered among classes (A+B). }
\label{fig:exam}
\end{figure*}
 
\section{Comparison of our models with two quasar spectra}\label{sec:comp}

In Fig.\,6a and 6b, we present our absorbed powerlaw fit to the class
(A) quasar PG\,1008+1319 and to the class (B) quasar Pks\,1229$-$02,
respectively.  In the case of PG\,1008+1319, it is interesting to
notice that the UV break is followed by a far-UV rise, which is well
modeled by our nanodiamond dust model, using \NN=1.2 and a dust
composition consisting of 70\% of Allende nanodiamonds and 30\% of
cubic nanodiamonds (\fdi=0.30).  The assumed spectral index is
$\anuv=+0.13$, a value determined by Neugebauer et\,al.  (1987). In
the case of the class (B) quasar Pks\,1229$-$02, the model shown in
Fig.\,6b assumes $\anuv=-0.4$ and is characterized by a column \NN=3.6
and dust composed mostly of cubic diamonds (D1) with \fdi=0.90. 

It appears that the six class (B) quasars in our sample are not only
more absorbed than class (A) but are extincted by dust  dominated by
cubic diamond grains (D1). On the other hand, the UV break in class (A)
spectra generally requires a mixture of the two dust flavors. The
absorption column is lesser in class (A) spectra.  The break in 39
class (A) quasars is well reproduced using a column in the range $0.6
\le \NN
\le 1.4$.

\acknowledgements 
 The authors acknowledge support from CONACyT grant 40096-F. 
We thank Randal Telfer for sharing the reduced HST FOS spectra.
Diethild Starkmeth helped us with proofreading.


\begin{thebibliography} 
  



\bibitem[Binette et al.(2005a)]{bm05a} Binette, L.,  Magris C., G., Krongold, Y.,
Morisset, C., Haro-Corzo, S., de Diego, J. A., Mutschke, H. \& Andersen, A., 2005a,  
\apj, 631, 661 (BM05)

\bibitem[Binette et al.(2005b)]{bi05b} Binette, L., Morisset, C., \& Haro-Corzo, S., 2005b,
in proc. of {\it The ninth
Texas-Mexico Conference on Astrophysics},  April
13--16, 2005, in San Antonio, Texas, eds. S. Torres \& G. MacAlpine,  \RMAASC, 23, 77

\bibitem[Binette et al.(2006)]{bm05c} Binette, L., Mutschke, H., Andersen, A., Haro-Corzo, S., 2006, 
in Proc. of 
``Granada Workshop on High Redshift Radio Galaxies'', Granada, 
18--20 April 2005, Ed. M. Villar-Mart\'\i n, E. P\'erez, R. Gonz\'alez-Delgado and J.L. G\'omez, 
Astronomische Nachrichten, 327, 151

\bibitem[Bohren \& Huffman(1983)]{bohren} Bohren, C. F. \& Huffman, D.R. 1983,
Absorption by Small Particles (New York: Wiley)


\bibitem[Edwards \& Philipp(1985)]{edwards} Edwards, D. F. \& Philipp, H. R. 1985,
in Handbook of Optical Constants of Solids, ed. E. D. Palik (Orlando:
Academic Press), 665
 



 








\bibitem[Mutschke et al.(2004)]{mutschke} Mutschke, H., 
Andersen, A.~C., J{\" a}ger, C., Henning, T., \& Braatz, A.\ 2004, \aap,
423, 983 


\bibitem[Neugebauer et al.(1987)]{neugebauer} Neugebauer, G., 
Green, R.~F., Matthews, K., Schmidt, M., Soifer, B.~T., \& Bennett, J.\ 
1987, \apjs, 63, 615 




\bibitem[Scott et al.(2004)]{scott} Scott, J., Kriss, G. A., Brotherton, M. S., Green, R. F.,  Hutchings, J., Shull, J. M. \& Zheng, W. 2004, \apj, 615, 135

\bibitem[Shang et al.(2004)]{shang} Shang, Z., Brotherton, M. S., Green, R. F., Kriss, G. A., Scott, J., Quijano, J. K., Blaes, O., Hubeny, I., Hutchings, J., Kaiser, M. E., Koratkar, A., Oegerle, W., \& Zheng, W. 2005, \apj,  619, 41

\bibitem[Telfer et al.(2002)]{telfer} Telfer, R. C., Zheng, W., Kriss, G. A., 
 \& Davidsen, A. F. 2002, \apj, 565, 773 (TZ02)
 

\bibitem[Van Kerckhoven et al.(2002)]{van02} Van Kerckhoven, 
C., Tielens, A.~G.~G.~M., \& Waelkens, C.\ 2002, \aap, 384, 568 



\end{thebibliography}
\end{document}